\documentclass[]{pasj01}

\let\leftroot\undefined
\let\uproot\undefined

\usepackage{amsmath}
\usepackage{newtxtext,newtxmath}
\usepackage[OT2,T1]{fontenc}
\usepackage{CJKutf8}

\usepackage{graphicx}	
\usepackage{color}
\usepackage[dvipsnames]{xcolor}
\usepackage{booktabs}
\usepackage{tcolorbox}
\usepackage{url}
\usepackage[implicit=False]{hyperref}

\newcommand{\Imcom}[1]{{\sc Imcom}}
\newcommand{\Treecor}[1]{{\sc Treecor}}
\newcommand{\Metacalibration}[1]{{\sc Metacalibration}}

\Received{$\langle$reception date$\rangle$}
\Accepted{$\langle$accept date$\rangle$}
\Published{$\langle$publication date$\rangle$}

\begin{document}

\title{Removing correlated noise stripes from the {\textit{\textbf{Nancy Grace Roman Space Telescope}}} survey images}
\author{Katherine Laliotis$^{* 1,2}$, Christopher M. Hirata$^{1,2,3}$, Emily Macbeth$^{1,2,3,4}$, \\
Kaili Cao ( \begin{CJK*}{UTF8}{gbsn}曹开力\end{CJK*}$\!\!$ )$^{1,2}$ }%
\altaffiltext{}{$^1$Department of Physics, The Ohio State University, 191 West Woodruff Avenue, Columbus, OH 43210, USA\\
$^2$Center for Cosmology and Astroparticle Physics, The Ohio State University, 191 West Woodruff Avenue, Columbus, OH 43210, USA\\
$^3$Department of Astronomy, The Ohio State University, 140 West 18th Avenue, Columbus, OH 43210, USA\\
$^4$Department of Astronomy, Steward Observatory, University of Arizona, 933 North Cherry Avenue, Tucson, AZ 85721, USA}
\email{laliotis.2@osu.edu}


\maketitle

\begin{abstract}
Weak gravitational lensing has emerged as a powerful tool for investigating the matter distribution in the Universe and how it has evolved over cosmic time. The Wide Field Instrument (WFI) on the \textit{Nancy Grace Roman Space Telescope} (\textit{Roman}) will deliver some of the highest precision measurements of weak lensing ever made. Since weak lensing is based on statistics of faint sources, it can be biased by even tiny instrument systematics, including correlated read noise. Previous works have shown the infrared detectors used in the \textit{Roman} WFI show correlations in their noise fields at a level significant for weak lensing measurements, even after application of standard reference pixel corrections; of particular concern is $1/f$ noise, which appears as horizontal banding in the detector frame. In this paper, we present \texttt{imDestripe}: a new \texttt{Python} module utilizing the multiple roll angles in \textit{Roman}'s observing strategy and linear algebra techniques to remove correlated noise stripes from observed images. We test \texttt{imDestripe} in a hybrid simulation by combining real noise realizations (from darks taken during ground testing) with simulated images of the astronomical scene, and find that the power spectrum of the banding can be suppressed by factors of 10--30 on large scales. We briefly discuss plans for further development of \texttt{imDestripe} in the context of the WFI pipeline.

\end{abstract}

\section{Introduction}

The accelerating expansion of the Universe is one of the major mysteries in modern cosmology. One approach to studying it is via the cosmic distance scale: this is how cosmic acceleration was first discovered (via the Type Ia supernova distance-redshift relation; \cite{Riess1998, Perlmutter1999}), and it continues to be an area of active research today \citep{Scolnic2022, DES2024, Rubin2025, DESI2025a, DESI2025b}. A complementary approach is to measure the growth of cosmological perturbations, which are sensitive to dynamical properties of dark energy or modifications to gravity. Weak lensing is an extremely promising, though extremely difficult, method for measuring the distribution of matter; in combination with other probes, precise and accurate weak lensing measurements will allow us to test fundamental aspects of our current standard model for the universe, including the nature of dark matter and dark energy (see, e.g., reviews by \cite{Hoekstra2008, Weinberg2013, Bartelmann2017, Prat2025}).

However, it is the specification of ``precise and accurate'' weak lensing measurements where the difficulty lies. ``Precise'' weak lensing requires a survey to be both deep and wide, to measure extremely large number of galaxy shapes across a range of redshifts. ``Accurate'' weak lensing requires not only very sensitive instruments but also control over sub-percent systematic effects that might introduce errors into the shape measurements themselves (\cite{Mandelbaum2018}). The previous generation of weak lensing surveys --- the Dark Energy Survey \cite{DES2022}, the Hyper Suprime Cam \cite{Miyatake2023, Sugiyama2023}, and the Kilo Degree Survey \cite{Wright2025} --- have reached total uncertainties on the amplitude of cosmic structure of a few percent. The next generation of surveys will aim for both larger samples and tighter control of systematic errors: these include \textit{Euclid} \citep{Euclid2025}, the Vera Rubin Observatory \citep{Ivezic2019}, and the \textit{Nancy Grace Roman Space Telescope} (\textit{Roman}, planned to launch as early as in 2026\footnote{\url{https://roman.gsfc.nasa.gov/}}). {\slshape Roman}'s Wide Field Instrument (WFI) will carry out an unprecedented imaging survey of the near-infrared sky \citep{ROTAC}. One of its central science goals is a thorough weak gravitational lensing survey \citep{Spergel2015, Akeson2019}. The Roman Science Requirements Document \footnote{\url{https://roman.gsfc.nasa.gov/science/docs/RST-SYS-REQ-0020D_DOORs_Export.pdf}} (SRD) specifies that galaxy shears must be measurable to within $\lesssim 5\times 10^{-4}$ root-mean-square systematic error per component to achieve its goals.

Near-infrared observations such as those with \textit{Roman} have several advantages for weak lensing, especially with a space telescope that is designed to observe from above the distorting effects of the atmosphere. Most of our source galaxies are redder than the zodiacal light. The longer wavelength means that for fixed sampling ratio\footnote{Defined by $Q = \lambda/(PD)$, where $\lambda$ is the wavelength; $D$ is the mirror diameter; and $P$ is the pixel scale projected on the sky. Fully sampled data has $Q\ge 2$; ``2$\times$'' undersampled data, which is commonly considered in algorithmic tests for weak lensing, has $Q\approx 1$. See, e.g., \citet{Kannawadi2021, Finner2023, Hirata2024} for more detailed discussions.}, we can achieve larger \'etendue per pixel. Photometric redshifts using the Balmer + 4000 \AA\ features over the whole redshift range of interest requires both visible and infrared data, an area where we expect the combination of Rubin + \textit{Roman} data to be particularly powerful. However, infrared detector technology is relatively new in weak lensing, and the non-ideal effects in these devices can be very different from charge-coupled devices (CCDs) used in previous visible-wavelength weak lensing surveys.

The \textit{Roman} detectors are part of the Teledyne HxRG family of imaging detector arrays \citep{Blank2011}. Since the \textit{Roman} detectors are new and have larger formats (4k$\times$4k) and different pixel design and process changes relative to previous infrared detectors \citep{Mosby2020}, it is essential to analyze laboratory and ground test data to prepare systematic mitigation strategies. Many groups have been using this data to measure potential sources of systematic error for the \textit{Roman} weak lensing survey. \citet{Givans2022} studied charge diffusion and quantum yield in Roman's H4RG detectors, and found that charge diffusion contributes a strong but correctable contamination to the weak lensing shear signal; quantum yield effects are not significant at the wavelengths of the weak lensing observations. \citet{Choi2020}, \citet{PlazasMalagon2024}, and \citet{Paine2025} have studied the brighter-fatter effect and its impact on the stars that will be used to measure the point-spread function. \citet{Freudenburg2020} explored the vertical trailing pixel effect (a cross-talk phenomenon in the readout circuit).

Yet another source of bias on shape measurements is correlated detector noise (see, e.g. \cite{Rauscher2011} for a discussion of correlated noise), which can introduce biased second moment measurements, biased galaxy selections, and complications in error propagation. \citet{Laliotis2024} showed that the level of correlated $1/f$ noise on \textit{Roman} SCAs is significant, and causes a bias on measured galaxy shapes that exceeds the requirement level from the SRD. 

While correction of biases within the weak lensing measurement pipeline or at catalog level may be possible, it is preferable to mitigate instrument-level systematic errors as early as possible in the process. To this end, removal of correlated instrument noise must be a significant step in the \textit{Roman} image processing pipeline. The first step in this mitigation is the reference pixel correction. \textit{Roman} detectors each possess a set of reference pixels, which do not collect photons from the sky, but are otherwise sampled and read in much the same way as the science pixels; and they contain a reference output, which does not read a physical pixel but traverses the same path from the sensor to the electronics where the voltages are digitized. The Improved Roman Reference Correction (IRRC; Rauscher et al, in prep.) will be applied to \textit{Roman} images during the exposure-level processing (conversion of raw data to 2-dimensional calibrated images) as a first level of noise removal. IRRC uses the reference pixels on each detector to optimally correct correlated noise on resultant frames using linear algebra techniques. However, even after application of IRRC some correlated noise persists at levels above the theoretical limit; correlated noise structures still appear in the up-the-ramp slope images \citep{Betti2024}.

To further correct the remaining $1/f$ noise, in this work we present \texttt{imDestripe}: a de-striping algorithm to remove correlated noise stripes from \textit{Roman} images. The algorithm is modeled off of similar ideas developed in the Cosmic Microwave Background (CMB) scientific community \citep{Delabrouille1998, Keihanen2004, Sutton2009}. In essence, the method of destriping works by combining multiple rolled exposures of a single region of sky to effectively scramble the coherent patterns present in the background. The difference between a single exposure and its interpolation is fit to a model whose parameters are optimized to minimize the difference between the image and its interpolation. The model converges to best-fit values representing the noise stripes.

In adapting the destriping method to our needs, we considered several fundamental complications that might arise for \textit{Roman} images specifically:
\begin{enumerate}
    \item Accounting for a much larger data volume. Esch \textit{Roman} exposure if $4096\times4224$ pixels, and we plan to process a full \textit{Roman} mosaic image (a $1\times1$deg$^2$ region composed from $\approx500$ images), at once; the destriping model has $\sim 2$ million parameters.
    \item Implementation of bright object and cosmic ray masks.
    \item Assessing the impact of image undersampling and a more complicated PSF, which factor into the projection matrix. 
\end{enumerate}

While the data volume and object masks did impact our implementation of destriping for this work, we found that the intrinsic undersampling and complex PSF of \textit{Roman} images did not require any specific treatment.

The structure of this paper is as follows. In Section~\ref{sec:detectors}, we describe the detectors on the \textit{Roman} Wide Field Instrument, their known noise properties, and the current status of noise corrections. 
Section~\ref{sec:imDestripe} outlines our implementation of noise destriping in \texttt{imDestripe}. We discuss the data used for this work and lay out the destriping process in detail step-by-step.
Tests of \texttt{imDestripe}'s performance are outlined and presented in Section~\ref{sec:performance}. Finally, we summarize and discuss results of this work in Section~\ref{sec:discussion}. A detailed discussion of the code implementation of \texttt{imDestripe} can be found in the Appendix.

\section{Correlated Noise on the \textit{Roman} H4RGs}\label{sec:detectors}

\subsection{H4RG detectors and noise}

The \textit{Roman} telescope Wide Field Instrument (WFI) is made up of 18 H4RG-10 Sensor Chip Assemblies (SCAs), which were manufactured by Teledyne. The ``10'' refers to the pixel pitch of 10 $\mu$m. The array dimensions are $4096\times 4096$ physical pixels. Each SCA has ``reference pixels,'' which are not light-sensitive and instead serve to measure the behavior of detector electronics. The reference pixels form a band 4 pixels wide on each edge of the detector, making the ``active'' region of an SCA $4088\times4088$ pixels. The SCAs generally consist of three layers: an anti-reflective (AR) coating, a Hg$_{1-x}$Cd$_x$Te semiconductor infrared absorber layer with $\approx 0.5$ eV band gap, and a silicon readout integrated circuit (ROIC) (as well as packaging). Each pixel acts as a photodiode, converting photons into electron-hole pairs, and connects to the ROIC via an indium ``bump'' (\cite{Mosby2020}). 

The voltage on each pixel is read non-destructively via a CMOS source follower; the pixels are read at 200 kHz, and are grouped into 32 readout channels (each 4096 rows $\times$ 128 columns). A full readout of the array takes $\approx 3.04$ s (this is longer than the ideal 2.62 s due to overheads, as well as gaps when the guide window is being used). These 32 channels, as well as a 33rd ``reference output'' that does not sense a physical pixel, are sent as analog data along a flexible cable connecting the SCA to the ACADIA controller (which provides bias and clock signals to the SCA, and which also contains the analog-to-digital converters where each sample of the analog voltage is converted into a 16-bit integer; see \cite{Loose2018}).

\begin{figure}
   \centering
   \includegraphics[width=0.9\linewidth]{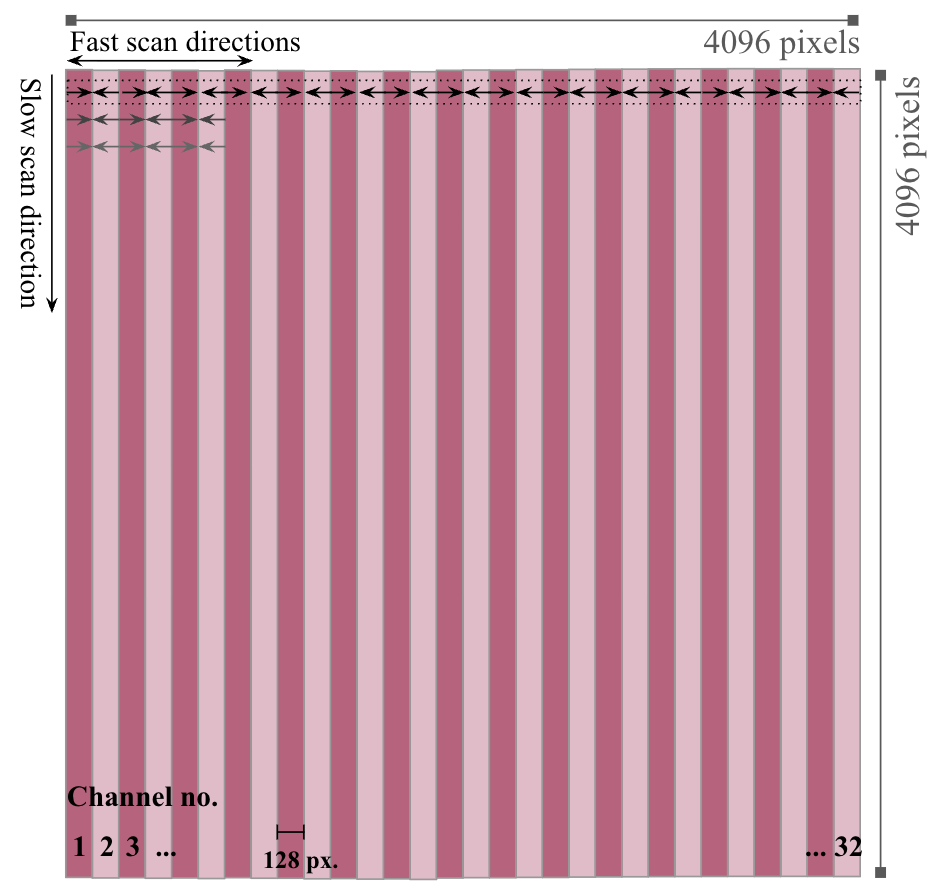}
   \caption{An illustration of a \textit{Roman} SCA layout. Pixels that are read out at the same time in the fast-scan direction show a read noise correlation.}
   \label{fig:Readout}
\end{figure}

An illustration of a \textit{Roman} SCA and its readout pattern is shown in Figure \ref{fig:Readout}. Pixel rows are read out one at a time, while for each row, all channels are read out simultaneously. The $128$ pixels in each channel are read out in sequence, with the order alternating for each channel to reduce correlations. As will be discussed further in this paper, dark exposures of \textit{Roman} SCAs show correlated noise with a \textit{1/f} power law, which manifests as horizontal bands across pixel rows in the image. 

During development of the H4RG-10 detectors, the pixel design was optimized to minimize the total noise, along with other detector systematics (\cite{Mosby2020}). These detectors typically show both white noise and pink, or $1/f$, read noise. To characterize the performance of the \textit{Roman} detectors, multiple rounds of testing were done by the team at the Detector Characterization Laboratory (DCL) at the NASA Goddard Space Flight Center. For this work, we use a set of images acquired in the Focal Plane Test in April 2023, when the 18 SCAs, flexible cables, and ACADIA controllers had been integrated into the flight Focal Plane System.

Data rate limits prevent the possibility of downlinking every frame from an exposure; instead, groups of frames are averaged together through a ``multi-accumulation'' (MA) process into $\approx8$ ``resultant'' frames which are transmitted from the spacecraft down to Earth for processing at Space Telescope Science Institute (STScI). The precise MA choices will be different for each of \textit{Roman}'s Core Community Surveys (CCS). The final MA configurations have yet to be determined, but several different options are currently being tested for optimal information content (and minimal noise) for the different science cases prioritized by each CCS. 

\begin{figure}
    \centering
    \includegraphics[width=0.9\linewidth]{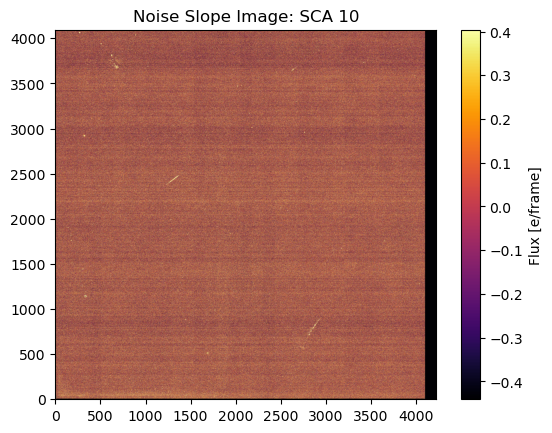}
    \caption{An example slope image made from a dark integration on SCA 10. Hot pixels, characteristic \textit{1/f} noise stripes, and the reference output channel (the dark, rightmost channel) can be seen here.}
    \label{fig:slopeimg}
\end{figure}

Once the MA frames have been downlinked, each pixel's signal over frames (recall \textit{Roman} images will be read out nondestructively) is linearly fit once per exposure. The slope of this line becomes the signal per pixel in the final images, producing a ``count-rate'' or ``slope'' image. An illustration of this process can be found in Figure 2 of \cite{Laliotis2024}. These slope images are the ``Level 2'' data products that will be directly accessible to public users of \textit{Roman} data\footnote{See the Roman Documentation \href{https://roman-docs.stsci.edu/}{Rdox} for more information about Roman data products.}. An example dark slope image is shown in the upper left panel of Figure \ref{fig:slopeimg}. The overall noise level is small, but clearly nonzero and containing some interesting features. 

\begin{figure*}
    \centering
    \includegraphics[width=0.99\linewidth]{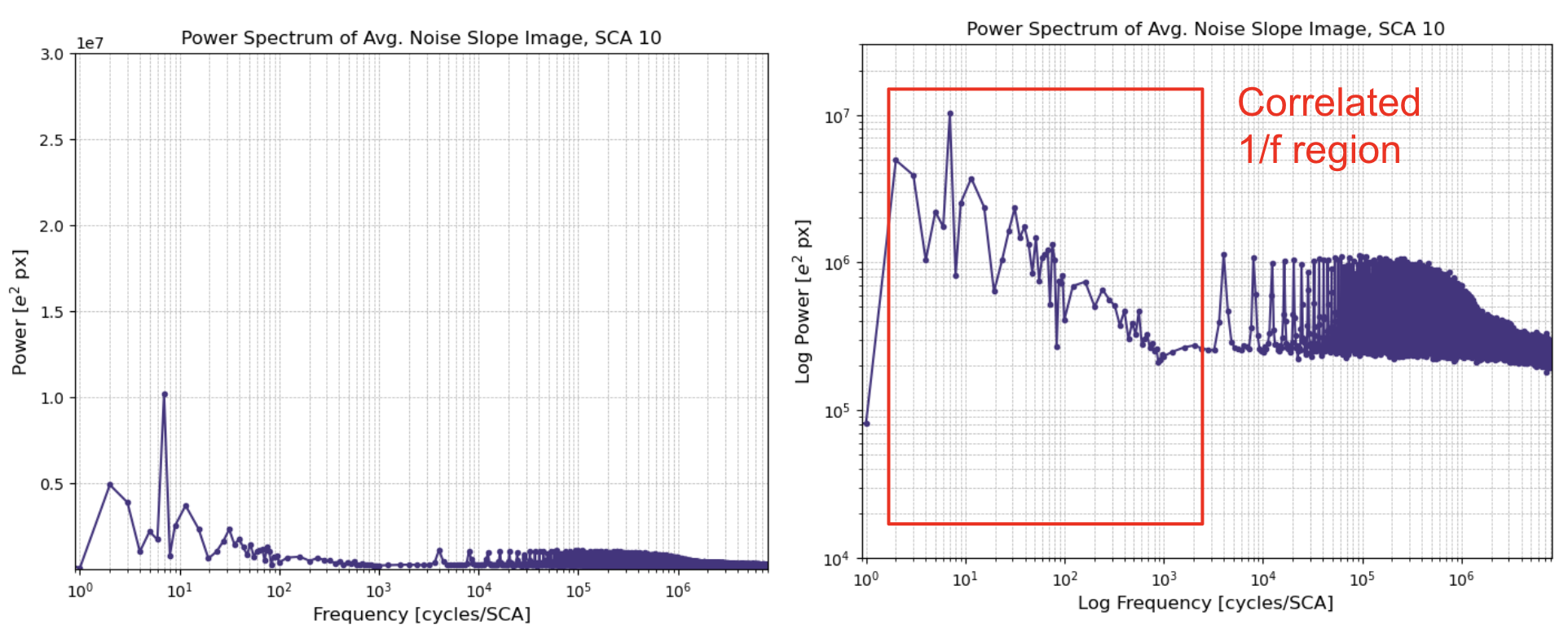}
    \caption{Binned power spectrum of the average noise slope image for SCA 10. The frequency axis shows the cycles each mode completes across the image length-- note the peaks at 4K and 8K. \textit{Left:} Linear scale power spectrum shows stationary noise on most scales. \textit{Right:} Log-log plot reveals a region of correlated $1/f$ noise at low frequencies.}
    \label{fig:sca10_noise}
\end{figure*}

The majority of the noise on the \textit{Roman} H4RGs is stationary white noise. Figure \ref{fig:sca10_noise} illustrates this by plotting the power spectrum of the noise from an average of 52 dark slope images for \textit{Roman} flight SCA 10. We compute the power spectrum on the dark slope image via \texttt{Scipy} implementation of the Fast Fourier Transform algorithm \citep{Scipy} after rescaling it to units of electrons per exposure, leaving the final power spectrum in $e^2$ pixels. The FFT measures power in frequencies spanning the image, in this case the width of the SCA. The frequency axis measures a number of cycles each mode completes across the length of an image.

The left panel shows clearly that most frequencies have constant power, meaning that the noise is uncorrelated and stationary. Plotting on a log-log scale (right panel of Figure~\ref{fig:sca10_noise}), we see a distinct slope in the power spectrum in low frequencies, suggesting the noise is correlated in this range. 

 For this paper, we focus on the correlated $1/f$ noise, so we will briefly describe the origin of that aspect of noise only. The causes of various types of noise in HXRG detectors in general are described in depth in \cite{Rauscher2011}. 
 



After pixels are read out via their individual source-followers, their signal is transferred to a column bus. Finally, the signal is fed from the column bus to an output source-follower, which is expected to cause another set of $1/f$ noise. This is believed to be the $1/f$ noise we observe in the output darks. Because the noise is generated in the readout electronics, it is different for each exposure, yet independent of the actual sky image being taken.

\subsection{Current state of correlated noise correction}

While white noise is uncorrelated among the various pixels, $1/f$ noise is highly correlated along the ``horizontal'' direction in the image, which can lead to biases in measurements one may want to take using the WFI. \cite{Laliotis2024} showed that correlated noise at the level shown in \textit{Roman}'s detectors induces significant additive bias on galaxy shape measurements. The estimated bias was above the required level for the weak lensing survey\footnote{See: \href{https://roman.gsfc.nasa.gov/science/docs/RST-SYS-REQ-0020D_DOORs_Export.pdf}{Roman Science Requirements Document}} by between 1 and 3 orders of magnitude depending on the observation band, illustrating a need for a method of correcting correlated noise if weak lensing measurements are to reach their required precision.

There are several existing methods of correcting for noise stripes on images. In \cite{Laliotis2024}, a simple correction scheme is applied using reference pixels as anchor points to a linear fit to the dark signal across each row. The value of the noise signal at each pixel position is then subtracted from the pixel value in the image. Our result demonstrated that this simple reference pixel correction scheme is insufficient for \textit{Roman}-level precision.

The \textit{Roman} image processing pipeline will use the Improved Roman Reference Correction (IRRC) algorithm from Rauscher et al, in prep. for reference pixel correction. This algorithm, adapted from the Simple Improved Reference Subtraction (SIRS; \citet{Rauscher2022}) used for JWST NIRCam images, utilizes linear algebra methods to optimally reduce correlated noise. \citet{Betti2024} demonstrate that the IRRC method is superior to simple reference row subtraction and reduces power in $1/f$ noise by a factor of 20 in individual resultant frames. However correlated noise is still not completely removed; the power spectrum of an up-the-ramp slope image constructed from IRRC-corrected resultant frames shows only marginal reduction in the low frequency $1/f$ noise power, compared to a slope image with no correction applied (see Figures 8-9, \citet{Betti2024}). 

\textit{Roman} operates in a new regime of frequency space, due to its small pixels, continuous readout strategy, and relatively short ($\approx150$s) exposure times. Because of this, the spatial variation of Roman's sky backgrounds will be \textit{dominated} by $1/f$ noise if this is not mitigated. It is thus essential that correlated noise be corrected to a high level in order to enable precision science with \textit{Roman}. 
The best path forward towards minimizing correlated noise for \textit{Roman} will be a combination of previously used methods and new ones. The image processing pipeline will employ a dual strategy of correlated noise mitigations, using both reference pixel subtraction through IRRC applied on resultant frames and through destriping via \texttt{imDestripe} applied to slope images prior to image coaddition. 

\section{imDestripe}
\label{sec:imDestripe}

\begin{figure*}[htb]
    \centering
    \includegraphics[width=0.8\linewidth]{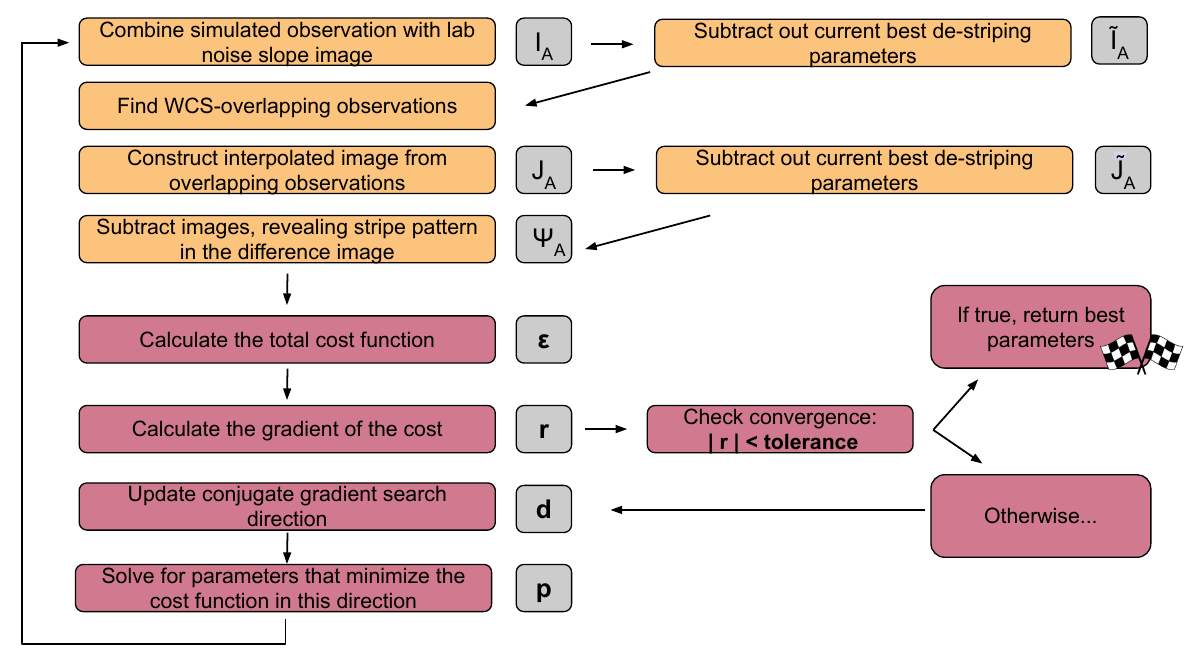}
    \caption{Flow chart depicting the main process of \texttt{imDestripe}. Light blue directives are calculated image-by-image for each SCA contributing to the mosaic, while the dark blue steps are executed while considering all SCAs at once. The purple squares show the variables representing the product of each step, to help the reader relate steps described here to equations in the paper.}
    \label{fig:flowchart}
\end{figure*}

We now describe the current version of \texttt{imDestripe}, our method of removing correlated noise from IR detectors. \texttt{imDestripe} will be integrated into the High Latitude Imaging Survey (HLIS) Project Infrastructure Team (PIT)'s image processing pipeline for \textit{Roman} cosmological analysis. The input to this pipeline will be Level 2 data products, which have already undergone calibration (including reference pixel correction and IRRC). After applying additional calibrations such as \texttt{imDestripe} from the PIT pipeline, images will be coadded with \textsc{PyIMCOM}; the resulting data products (including striping models as well as the final co-added images) will also be made publicly available.

The \texttt{imDestripe} algorithm is designed to be run on a single mosaic, which is made up of $\sim 500$ individual SCA images $I_A$ and covers $\sim 1$ deg$^2$ of the sky. Overall, the algorithm is designed to remove stripes of correlated noise from the input images, ultimately solving for an optimally flat sky background in the mosaic image. 

In general, we will use capital Roman letters $A, B, ...$ to refer to SCA IDs on the focal plane. For a single exposure, $A$ could only range from 1--18, but since we will be handling multiple exposures at once, the index $A$ will actually have a much larger range due to the large number of exposures. Rows of pixels will be denoted by lowercase Greek letters, $\alpha, \beta,...$, which will range from 0--4087, as we consider only the active pixels. Finally, individual pixels are indexed by lowercase Roman letters $i, j, ...$, which will range from $0\le i, j \le 4088^2-1$. The signal in pixel $i$ of image $A$ would thus be denoted $x_{Ai}$. 

We will also introduce the symbol $\delta_{\alpha i}$ such that:
\begin{equation}
\delta_{\alpha\,,Ai} = \begin{cases}
1 & \text{if pixel } i \text{ is in row } \alpha \text{ of image A},\\
0  & \text{if pixel } i \text{ is not in row } \alpha \text{ of image A.}
\end{cases}
\end{equation}

A flow chart representing the main steps of the process can be found in Figure \ref{fig:flowchart}.


\subsection{Simulation setup}

We test {\tt imDestripe} on a set of simulations. Here the input images are taken from the \citet{OpenUniverse2025} simulation of the \textit{Roman} High-Latitude Wide Area Survey. We selected the set of $482$ images that were processed via {\tt PyIMCOM} as a representative ``mosaic'' image. A mosaic image covers a $\sim1\times1$ deg$^2$ region of sky, and describes the set of images processed through {\tt PyIMCOM} in one run (see Figure 4 and the related discussion in \citet{Hirata2024}). We use the ``simple'' simulations, which represent a calibrated exposure-- relevant sky backgrounds and noise sources are present, but detector effects are not included. In the SOC data structure, these images simulate ``Level 2'' data products.

First, we import a given image $I_A$ which is inside the mosaic. We add in real detector noise images $N_A$, which are slope images of read noise frames collected in lab experiments plus a Poisson sky background (see \cite{Laliotis2024} for processing details). The image is then multiplied by a hot pixel mask derived from lab data, $M_A$ to remove degenerate pixel regions from the image. 

\subsection{Masking}

The first step in \texttt{imDestripe} is to mask out bright objects (e.g., bright stars, cosmic rays, etc.), so that their large flux does not bias the background calculation. For this work, we choose to mask out pixels $I_{Ai}$ with flux F greater than 2.5 times the median pixel value in the pre-sky-subtraction image, along with the eight pixel nearest neighbors to avoid contamination. The threshold for bright object masking is set in the configuration file where the user may specify an additive factor $c$ and multiplicative factor $m$ to mask pixels above the value of $m\times \text{Med}(I_{A})+c$.


\subsection{Correlated noise model}

We model stripes of correlated noise in images by a vector ${\boldsymbol p}$ of constant offset parameters, with one parameter for each row $\alpha$ in each image:
\begin{equation}\label{eq:model}
    N_{Ai} = \sum_{\alpha=0}^{4088-1} p_{A\alpha}\delta_{\alpha, Ai}
\end{equation}
where $N_{Ai}$ is the noise image for SCA exposure A. For the full mosaic containing $N_{\rm im}$ images, ${\boldsymbol p}$ is a vector of length $4088\times N_{\rm im}$. Throughout the \texttt{imDestripe} process we will utilize both this ``flattened'' representation and an image representation of ${\boldsymbol p}$ as an $(N_{\rm im}, 4088, 4088)$ data cube.

Each \textit{Roman} SCA is approximately $\frac18$ deg wide, and mosaic images are $1\deg\times1\deg$, so it takes $\sim 64$ exposures to cover a mosaic image. For ideal coverage of a region of sky, we aim for 6 individual exposures of each region. This results in ${\boldsymbol p}$ of length $\approx 2\times10^6$. We leave open the option to include other parameters (such as sky background level) to be fit to the mosaic, so ${\boldsymbol p}$ may in principle be larger than this estimate.

We use ${\boldsymbol p}$ to create a de-striped image: \begin{equation} \label{eq:IAi}
    \tilde{I}_{Ai} = I_{Ai} - \sum_{\alpha} p_{A\alpha} \delta_{\alpha\,,Ai} 
\end{equation}
where $A$ is a single image, $i$ denotes a single pixel (with coordinates $(x_i, y_i)$ ), and $\alpha$ denotes a particular row of pixels.

\subsection{Cost function}

Before solving for the optimal parameters ${\boldsymbol p}$ that will accurately remove our image's correlated noise background, we need a method for testing the accuracy of those parameters. We define a cost function for the de-striped image:
\begin{equation}
\epsilon_A(\mathbf{p_{A}}|\tilde{I}_{Ai}) = \sum_{i} f \bigl( \tilde{I}_{Ai}-\tilde{J}_{Ai} \bigr) .
\label{eq:epsA} 
\end{equation}
Here $f$ is an error metric; the most common choice, which we adopt for this work, is the sum of squares metric, $f(x)=x^2$. We also define $\tilde{J}_{Ai}$ as an interpolated version of $\tilde{I}_A$ constructed from other de-striped images ($\tilde{I}_B, \tilde{I}_C, ...$) that overlap image A:
\begin{equation}
\tilde{J}_{A, i} = \frac{\sum_{B,j} G_{Bj} w_{Ai;Bj}\tilde I_{Bj}}{G_{Ai}{\cal N}_{B|Ai}} ,
\label{eq:JAi}
\end{equation}
where:
\begin{itemize}
    \item the sum is over pixels $j$ in image $B$ that overlap pixel $A, i$;
    \item $w_{Ai;Bj}$ are interpolation weights, with $\sum_j w_{Ai;Bj}=1$; and
    \item ${\cal N}_{B|Ai}$ is the number of overlapping images $B$ contributing to the sum.
\end{itemize}
Further details on how the interpolation is calculated can be found in the Appendix. 

The interpolated image is also rescaled to account for differences in the pixel response function between pixel $j$ in image $B$ and pixel $i$ in image $A$. We do this by calculating an effective gain:
\begin{equation}
    G_{Bj} \propto \Bigl[ t_{\text{exp},B} \Omega_{Bj} \int \frac{d\lambda}{\lambda} A_{\text{eff},Bj}(\lambda)\Bigr]^{-1}.
\label{eq:Geff}
\end{equation}

For the current simulated images from \citet{OpenUniverse2025}, the effective area $A_{\rm eff}$ integral and the exposure time $t\rm{exp}$  are both constant in each image (though this will have to change in the next simulation when the flat field is incorporated). The pixel areas $\Omega_{Bj}$ vary due to distortions. Effective area integrals in each observation filter are calculated in \cite{Laliotis2024}. 

The total error in a mosaic image is then calculated as $\epsilon(\mathbf{p}) = \sum_{A} \epsilon_A\,(\mathbf{p_A}|I_{Ai})$.

\begin{figure*}[htb]
    \centering
    \includegraphics[width=0.95\linewidth]{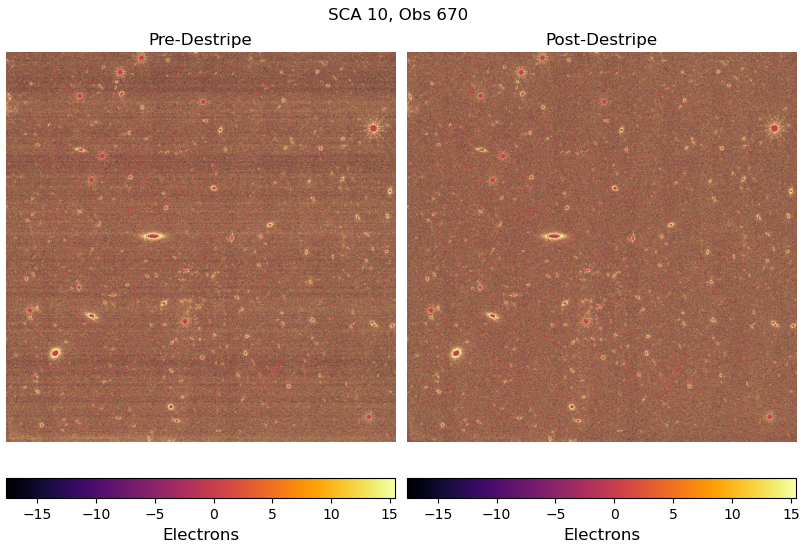}
    \caption{Comparison of a masked simulated image (Observation 670 on SCA 10) before and after destriping. \textit{Left:} The masked image before destriping; horizontal stripes can be seen in the noise. \textit{Center:} The masked image after destriping; horizontal stripes are no longer visible. \textit{Right:} The optimal offset parameters for this image.}
    \label{fig:trip_real}
\end{figure*}

\subsection{Finding the best-fit de-striping parameters}

To solve for the optimal de-striping parameters ${\boldsymbol p}$, we utilize the method of conjugate gradient descent\footnote{We refer the interested reader to \href{https://www.cs.cmu.edu/~quake-papers/painless-conjugate-gradient.pdf}{An Introduction to the Conjugate Gradient Method Without the Agonizing Pain} for further details on the conjugate gradient algorithm and relevant choices within it}, minimizing the cost function (Eq.~\ref{eq:epsA}). We implemented both linear and non-linear conjugate gradient methods, but found that the quadratic error metric $f(x)=x^2$, which enables the much faster linear conjugate gradient method, works well, and so this is our current baseline. 

The direction of steepest descent is given by the negative of the residual, 
\begin{equation} \begin{split} \mathbf{r}
& =\nabla_p \epsilon(\mathbf{p}) \\ 
& =\sum_{A} \Biggl\{ \sum_{i} \Bigl[ f' \bigl( \tilde{I}_{Ai}-\tilde{J}_{Ai} \bigr) \bigl(\frac{\partial\,\tilde{I}_{Ai}}{\partial\,p_{A,\alpha}} -\frac{\partial\,\tilde{J}_{Ai}}{\partial\,p_{A,\alpha}} \bigr) \Bigr] \Biggr\} \\
& =\sum_{A} \Biggl\{ \sum_{i} \Bigl[ f' \bigl( \tilde{I}_{Ai}-\tilde{J}_{Ai} \bigr) \\
& ~~~\times \bigl(\frac{\partial\,\tilde{I}_{Ai}}{\partial\,p_{A,\alpha}} - \frac{\sum_{B,j} G_{Bj} w_{Ai;Bj} ( \partial\,\tilde{I}_{Bj}/\partial\,p_{A,\alpha} )}{G_{Ai} {\cal N}_{B|Ai}} \bigr) \Bigr] \Biggr\} \\
& = \sum_{A} \Biggl\{ \sum_{i}  f' \bigl( \tilde{I}_{Ai}-\tilde{J}_{Ai} \bigr) \biggl(-\delta_{\alpha,Ai}+\sum_{Bj}\frac{G_{Bj}w_{Ai;Bj}}{G_{Ai}{\cal N}_{B|Ai}}\delta_{\alpha,Bj}\biggr) \Biggr\}.
 \end{split}
 \end{equation}

The general conjugate gradient algorithm is an iterative method for solving large systems of linear equations. We outline the general steps below; our full implementation is detailed in Appendix~\ref{app:codenotes}. \texttt{imDestripe} is open-source and can be found in the \textsc{PyIMCOM} repository of the HLIS PIT GitHub Organization\footnote{\href{https://github.com/Roman-HLIS-Cosmology-PIT/pyimcom}{https://github.com/Roman-HLIS-Cosmology-PIT/pyimcom}}.

To start, the parameters ${\boldsymbol p}_{(0)}=0$. The first gradient is calculated ${\boldsymbol r}_{(0)}=-{\boldsymbol \nabla}\epsilon(I, J, {\boldsymbol p}_{(0)})$, using the cost function $\epsilon$ defined above and generally dependent on the image $I$, its interpolated version $J$, and the current de-striping parameters ${\boldsymbol p}$. After this initial iteration, the algorithm repeats the following steps for each iteration, denoted by the index $(n)$:
\begin{enumerate}
    \item Calculate the direction of steepest descent, 
    \begin{equation}
    {\boldsymbol r}_n = -{\boldsymbol \nabla}\epsilon(I, J, {\boldsymbol p}_n).
    \end{equation}
    \item Update the direction step size $\beta$ using the Polak-Ribi\`ere method,
    \begin{equation}
        \beta_{n} = \frac{{\boldsymbol r}_{n}^{\rm T}({\boldsymbol r}_{n}-{\boldsymbol r}_{n-1})}{{\boldsymbol r}^{\rm T}_{n-1} {\boldsymbol r}_{n-1}}.
    \end{equation}
    (We set $\beta_0=0$.)
    \item Update the conjugate direction, ${\boldsymbol d}_n = {\boldsymbol r}_n + \beta_n {\boldsymbol d}_{n-1}$.
    \item Using a linear search, minimize $\epsilon(I, J,{\boldsymbol p}_n + \alpha_n{\boldsymbol d}_n)$ to recover the optimal direction depth for the next step $\alpha_n$. 
    \item Update the parameters ${\boldsymbol p}_{n+1} = {\boldsymbol p}_n + \alpha_n{\boldsymbol d}_n$.
\end{enumerate}

Convergence can be determined by either a tolerance factor $C_{\rm tol}$ (the algorithm would converge when $||{\boldsymbol r}_n|| \leq C_{\rm tol}\,||{\boldsymbol r}_{(0)}||$) or by a set number of iterations. Because of the large number of parameters and intrinsic variation in the images means, the norm is large. We found that in general it drops by large amounts for the first several iterations and then starts to decline less steeply towards an eventual plateau. In this work we run 11 iterations of conjugate gradient, where the overall norm started to converge towards a steady value. 

Finally, the linear search is carried out by first calculating a test value of $\alpha$ based on the current gradient and then
calculating derivatives evaluated at the current and test values of $\alpha$. This secant-like update method approximates where the directional derivative would vanish, ensuring a descent step. 

The cost and gradient calculations run in parallel for each SCA in the mosaic, updating the relevant rows of the ${\boldsymbol p}$ vector as they go. The results of the cost and residual functions are summed up when needed to evaluate whether to converge. The decision to parallelize the algorithm was made in order to improve efficiency.

Once the conjugate gradient descent has converged on a set of parameters ${\boldsymbol p}_{\text{final}}$, we take the optimized destriping parameters and, subtracting them from their corresponding images, construct the final destriped images $\tilde I_{A, B, C...}$ which can then be coadded using \textsc{PyIMCOM} \citep{Cao2025} into a full mosaic image.

\section{Performance Tests}\label{sec:performance}

We ran tests of the \texttt{imDestripe} algorithm on the Cardinal cluster of the Ohio State Supercomputer. Each node has 2 Xeon CPU Max 9470 processors, with a total of 96 usable cores \citep{OSC2024}. On a test run with 482 overlapping exposures, using 1 node, the algorithm reached approximate convergence after $\sim15$ iterations of the conjugate gradient algorithm out of $20$ iterations over a total of $18.70$ hours. Each conjugate gradient iteration took $\sim53$ minutes. The job's memory usage oscillated around $45\pm15$ GiB of RAM. Figure~\ref{fig:ds_metrics} shows a summary of metrics describing the efficiency and evolution of the \texttt{imDestripe} test run.

The results of \texttt{imDestripe} can be seen qualitatively in Figure \ref{fig:trip_real}. The input image (left panel) has visible stripes of noise across it, and the final image (center panel) does not show the same pattern. Other noise background effects become visible once the overall stripes have been removed-- this will be discussed further in Section \ref{sec:discussion}. We also show the final offset parameters for this sample image.

To assess the success of \texttt{imDestripe} in removal of correlated noise, we compare several pre- and post- destriping noise statistics. We note here that the noise slope images used in this analysis have already been corrected by a simple reference pixel correction scheme described in Section 2.2 of \citet{Laliotis2024}. However, they have not undergone the Improved Roman Reference Subtraction (IRRC) algorithm, which we expect to be used as a first level of correlated noise mitigation on \textit{Roman} images; IRRC was not yet integrated into the SOC calibration pipeline at the time of this work \citep{Betti2024}. We plan to explore the total correlated noise mitigation achieved combining all current strategies we have in mind for \textit{Roman} in future work.

\begin{figure}[h]
    \centering
    \includegraphics[width=0.9\linewidth]{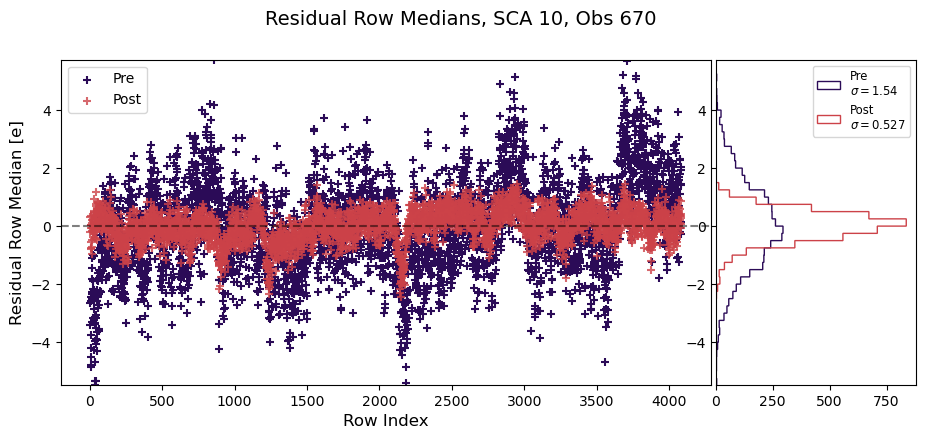}
    \caption{Row median residuals for SCA 10 in Observation 670, pre- (indigo) and post- (magenta) destriping.}
    \label{fig:rowresids}
\end{figure}

Following \citet{Rauscher2022} we define a noise residual to quantify the prominence of \textit{1/f} noise stripes in images. In a given image, we take the median value in each row. We then calculate the mean of that set, defining an overall median value for the image. The difference between each row's median and the overall median is what we define as the row median residual. This quantifies the overall scatter of the noise stripes in each image. Figure \ref{fig:rowresids} shows the noise residual for each row in Observation 670 on SCA 10, pre- (indigo) and post- (magenta) destriping. There is a clear decrease in the variance of the medians after \texttt{imDestripe} has been applied. This can also be seen in the histogram panel of the figure, where the medians after destriping are not only more strongly peaked at zero, but also more suppressed at large values, making the histogram appear closer to Gaussian. The standard deviation of the row median residuals goes from $1.53$ DN/frame before destriping to $0.55$ DN/frame after, a $3\times$ reduction.

\begin{table}[]
    \centering
    \begin{tabular}{c|c|c|c}
         \textbf{Image} & $\boldsymbol{\sigma}$ & \textbf{3rd Moment} & \textbf{4th Moment} \\
         \hline\hline
        Pre-Destripe & 1.543 & 0.311 & 18.415 \\
        Post-Destripe & 0.527 & -0.084 & 0.313 \\
    \end{tabular}
    \caption{Change in moments of the row residual distribution from application of image destriping to Observation 670, SCA 10-- these moments correspond to the histograms in the right panel of Figure \ref{fig:rowresids}}
    \label{tab:distributions}
\end{table}

Following \citet{Rauscher2022}, we further quantify the deviation of the correlated noise from Gaussian by calculating higher-order moments of our distribution. Deviation of these parameters from their expected values for a normal distribution would indicate that a model more complex than a constant offset (Eq.~\ref{eq:model}) might be needed to capture the correlated noise accurately. The values of the standard deviation, 3rd moment (skew), and 4th moment (kurtosis) of the row residual distributions for the example image used throughout this work (Obs. 670, SCA 10) are in Table~\ref{tab:distributions}. The standard deviation around the row medians decreases as discussed previously. The third and fourth moments also decrease in magnitude after image destriping.

We note that we do not expect the final destriped images to be perfectly Gaussian; the simulated images used in this work did contain different sky levels and other external noise sources. However, these backgrounds are subdominant to correlated read noise due to \textit{Roman}'s small pixels and short exposure time. Thus the presence of overall gradients or nonzero background levels in the final destriped images actually shows that \texttt{imDestripe} is able to optimally remove correlated noise for each individual exposure, without removing additional background flux not related to the stripes.

\begin{figure}[h]
    \centering
    \includegraphics[width=0.9\linewidth]{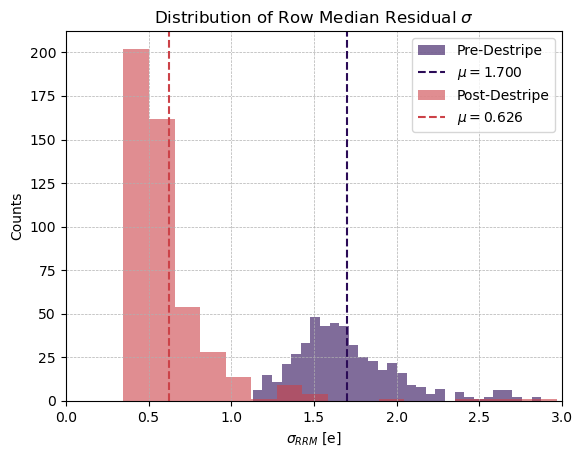}
    \caption{Distribution of standard deviation of residual row medians in all frames in this mosaic before and after application of \texttt{imDestripe}. The distribution and its mean both shift towards zero, illustrating the reduction in noise stripes.}
    \label{fig:rmr_stdev}
\end{figure}

We also calculate each exposure's mean standard deviation of row median residuals and present the total distribution of this statistic in all exposures in the mosaic before and after \texttt{imDestripe}. The pre-destriping average standard deviation of row means is $1.70$ e/exp in the whole mosaic. After destriping, this is reduced to $0.626$ e/exp. Figure \ref{fig:rmr_stdev} shows this overall shift towards a flatter background.

It is common practice to study the significance of image noise by investigating the noise power spectrum, the square magnitude of the image's Fourier Transform. Significant features of the power spectrum give us key insights about the contributing sources of noise. We use \texttt{SciPy}'s Fast Fourier Transform algorithm implementation \citep{Scipy} to compute the Fourier transforms required for this analysis.

\begin{figure}[htb]
    \centering
    \includegraphics[width=0.95\linewidth]{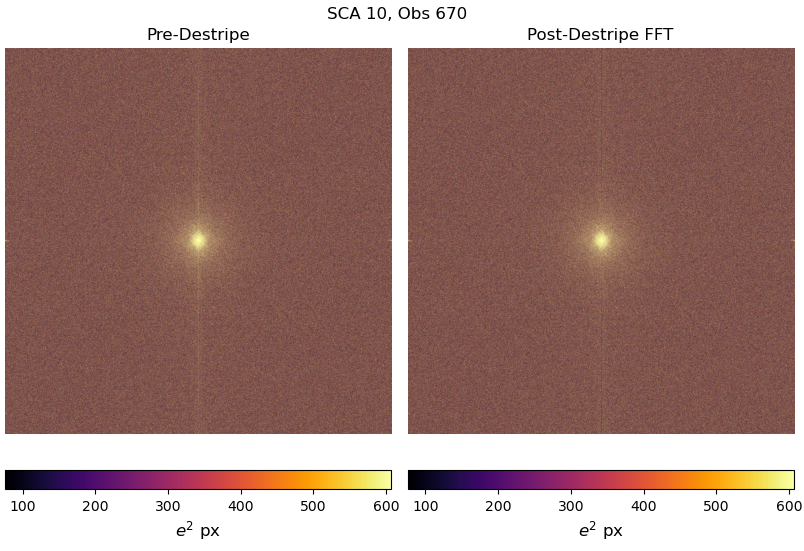}
    \caption{Power spectra of a masked simulated image (Observation 670 on SCA 10) before and after destriping. \textit{Left:} The masked image before destriping; the central bright vertical line is the FT of the horizontal stripe pattern. \textit{Right:} The masked image after destriping; the bright central feature has been significantly reduced. }
    \label{fig:trip_fourier}
\end{figure}

Figure \ref{fig:trip_fourier} shows the 2D power spectra of the simulated images before and after application of \texttt{imDestripe}. This image is a 2D FFT of Figure \ref{fig:trip_real}. The bright central line in the left panel of Figure \ref{fig:trip_fourier} corresponds to the horizontal noise stripes. After application of image destriping, the bright central line has been removed almost entirely. The total noise powers before and after the application of \texttt{imDestripe}, calculated by summing the 2D FFT power spectrum, are $6.406\times10^9 e^2/$px and $6.374\times10^9 e^2/$px; the power is reduced by $\sim3\times10^7 e^2/$px. 
The bright round feature in the center corresponds to Gaussian noise, and is expected to be present both before and after destriping.

\begin{figure}[h]
 \centering
    \includegraphics[width=0.9\linewidth]{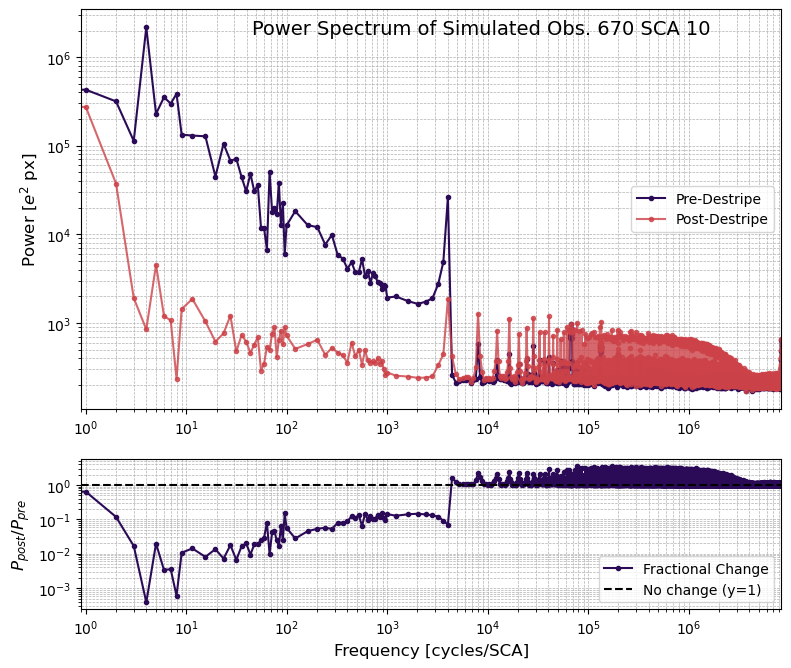}
    \caption{Power spectrum of a single up-the-ramp noise slope image (SCA 10) before (indigo) and after (magenta) processing with \texttt{imDestripe}. Suppression of noise in the correlated \textit{1/f} region is between one and two orders of magnitude.}
    \label{fig:PS_reduction}
\end{figure}

\begin{figure*}[t]
    \centering
    \includegraphics[width=0.99\linewidth]{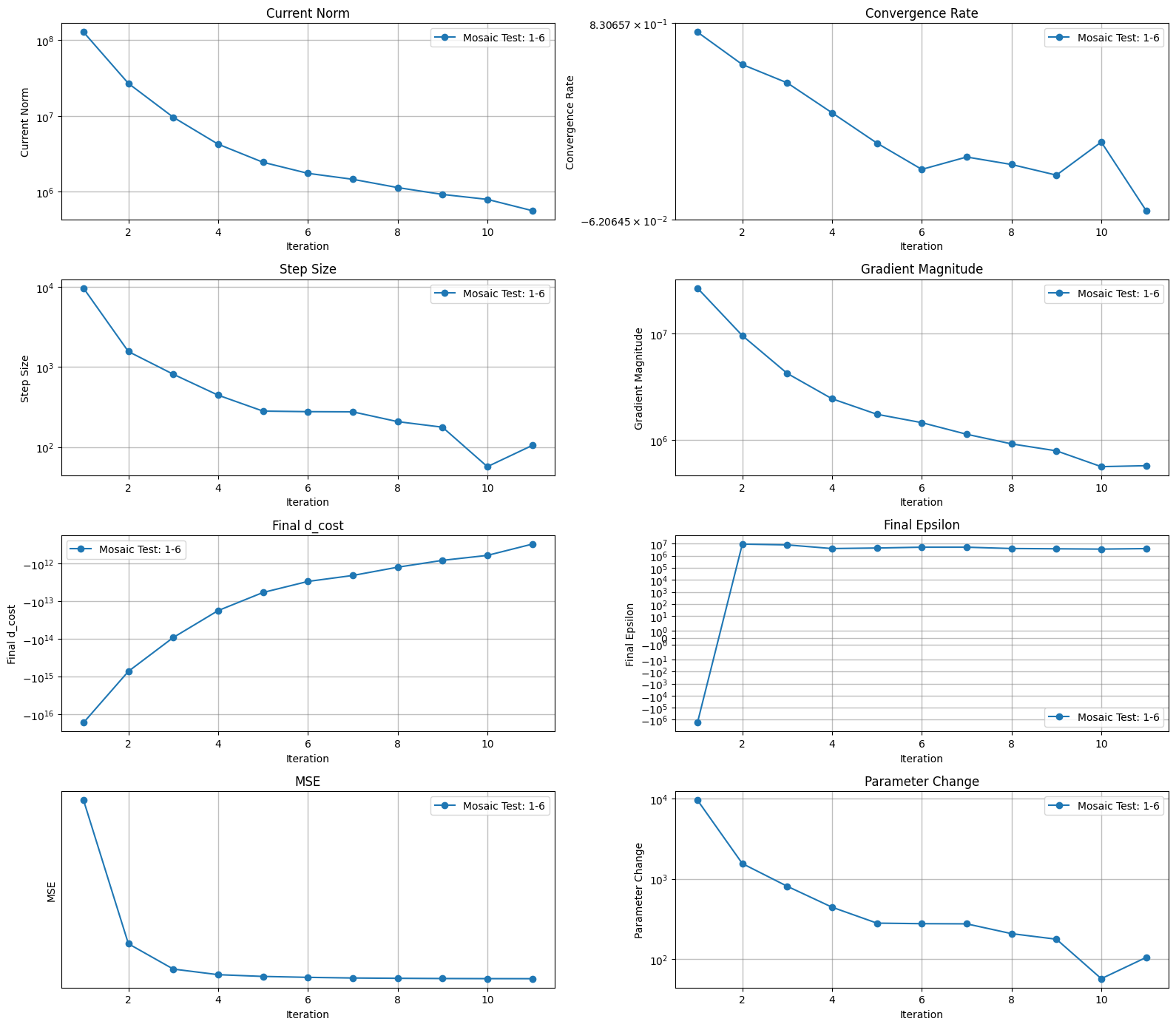}
    \caption{Metrics showing the computing performance of \texttt{imDestripe} as a function of CG iterations. Clockwise from upper left: Matrix norm of the current gradient, convergence rate-- defined as $(\|g_k\|-\|g_{k+1}\|)/\|g_k\|$, gradient magnitude $\|g_{k+1}\|$, total cost value $\epsilon_k$ (see Equation A8), change in parameter values quantified by $\|p_{k+1}-p_k\|$, and the mean squared error of $\Psi$ images, $\text{MSE}=\langle \Psi_{k+1}^2\rangle$. Note that iteration 1 here is after the initialization of parameters to zero, so it already includes some improvement from the initial values.}
    \label{fig:ds_metrics}
\end{figure*}

We also compute a flattened one-dimensional power spectrum of the up-the-ramp slope images before and after application of destriping, by raveling the arrays into 1D and again Fourier Transforming via \texttt{Scipy} FFT. The resulting spectra for one example image are shown in Figure~\ref{fig:PS_reduction}. The power spectrum of the initial noise image-- the noise slope image for one integration on SCA 10-- is plotted in indigo. The power spectrum of the final noise image-- the de-striped simulated image with the initial simulation subtracted off-- is shown in magenta. The reduction in the correlated $1/f$ region is clear; the lower panel of the figure shows that the amplitude of the correlated noise is suppressed by one to two orders of magnitude. Additionally, the noise is approximately stationary down to a frequencies of $\approx50$ cycles per exposure. We expect that this will decrease the impact of noise on later science measurements significantly.

\section{Summary and Discussion}
\label{sec:discussion}

The \textit{Roman} Space Telescope's High Latitude Wide Area Survey will provide the highest-resolution multi-band wide survey of the sky ever created. This is made possible by the Wide Field Instrument's 18 H4RGs, which have 0.11 arcsec pixels, a nondestructive readout mode, and excellent IR sensitivity. The \textit{Roman} images necessitates extreme control of all sources of systematic errors, especially for extreme-precision measurements like weak gravitational lensing --- a $\lesssim 1\%$ effect.

Correlated read noise is a known source of additive bias for weak lensing shear measurements, but previous weak lensing surveys have not operated in the regime where detector $1/f$ noise would be a prominent source of backgrounds.  We showed in \cite{Laliotis2024} that the correlated read noise from the \textit{Roman} detectors causes a bias in measured galaxy shapes that is not just significant but exceeds mission design requirements on all scales. 

In this paper, we have presented \texttt{imDestripe}: a new method to remove that correlated noise, and thus suppress the bias we would expect to see in the WL shape catalog. \texttt{imDestripe} uses \textit{Roman}'s multiple dithered and rolled imaging strategy to construct interpolated versions of each image in a mosaic and calculate the difference between each image and its interpolation. Then, it employs the method of conjugate gradient descent to optimally solve for a set of offset parameters describing the difference between the two images --- in this case, the parameters are in the form of a constant offset per pixel row in each image.

We have shown that {\tt imDestripe} reduces the power of correlated \textit{1/f} noise by a factor of 10--30 on the relevant frequency scales. 
The stripes are no longer able to be seen by eye in the final images, nor in the 2D power spectra of the images. Finally, we calculate the deviation of each row's median from the median background in the image. The scatter is overall reduced for each image, but coherent trends or drifts can be seen in the post-\texttt{imDestripe} row median residuals. These drifts are due to other sky background effects which are present in the simulations but not corrected for in \texttt{imDestripe}. 
In the future, we plan to expand the capacity of imDestripe to include vertical noise stripes from channel offsets and an overall sky background correction, accounting for overall signal drifts, variations in zodiacal light, and other usual sky background effects.

We note that the method shown in this work has a known issue with SCAs on the edges of the mosaic. If an SCA region has no overlap with other images in the mosaic, the solution can wander away and fit incorrect stripes to the image, resulting in a post-\texttt{imDestripe} image containing prominent stripes from an incorrect fit. The post-destriping histogram in Fig. \ref{fig:rmr_stdev} shows that a few images still have a large $\sigma_{\rm RRM}$ after destriping; these few outliers are the edge SCAs.

For image processing in the HLIS Cosmology PIT pipeline, it is likely that edge exposures will be excluded from analysis. Thus we are not concerned about the behavior of these edge images. In the future, we also plan to impose a prior on the parameter values. This will likely solve the edge image issue by preventing the wandering of the best fit parameters, in addition to improving the runtime of the algorithm.

Application of \texttt{imDestripe} to a combined dataset made of real detector noise and simulated \textit{Roman} images from \cite{OpenUniverse2025}
successfully reduced the level of correlated noise in the \textit{Roman} detector lab tests by one to two orders of magnitude on the relevant scales. We expect the implementation of this algorithm, along with the other planned correction schemes involving reference pixels, to significantly reduce the expected additive shear bias being caused by correlated noise. In the future, we plan to re-analyze the level of noise-induced bias at the pipeline level. Once \texttt{imDestripe} has been integrated into \textsc{PyIMCOM}, we will execute an end-to-end analysis of simulated \textit{Roman} observations incorporating all elements of correlated noise reduction (including IRRC \textit{and} destriping) and running image coaddition to assess the overall level of noise bias suppression.

\texttt{imDestripe} was designed as part of efforts by the Roman High Latitude Imaging Survey (HLIS) Project Infrastructure Team (PIT) to design infrastructure for measuring cosmological parameters with \textit{Roman} data. The PIT image processing pipeline will now include \texttt{imDestripe} as a processing step before images are passed to \textsc{PyIMCOM} for image coaddition. With \texttt{imDestripe}, we add to the arsenal of \textit{Roman} software serving not just the HLIS PIT but the astronomy community as a whole; \texttt{imDestripe} and all aspects of the PIT cosmology pipeline are open source code on GitHub, and the images and data products derived through application of this pipeline will be world public as well. 

\begin{ack}

 We thank Ami Choi and Arun Kannawadi for their continual support and helpful comments throughout this project.

 This work was supported by the NASA ROSES grant 22-ROMAN11-0011, contract number 80NM0024F0012, via a JPL subaward. 

 This material is also based upon work supported by the U.S. Department of Energy, Office of Science, Office of Workforce Development for Teachers and Scientists, Office of Science Graduate Student Research (SCGSR) program. The SCGSR program is administered by the Oak Ridge Institute for Science and Education for the DOE under contract number DE‐SC0014664.
 
 This paper is based on data acquired at the Detector Characterization Laboratory (DCL) at NASA Goddard Space Flight Center. We thank the DCL for making the data available to us for use in this project. 
 
Computations were performed on the Cardinal cluster at
the Ohio Supercomputer Center \citep{OSC}.
\end{ack}

\appendix

\section{Implementation notes}
\label{app:codenotes}

This appendix describes the implementation of \texttt{imDestripe}, and serves as a guide to the various functions. There are two sets of functions: those in the C layer of the code ({\tt furry-parakeet}\footnote{\url{https://github.com/Roman-HLIS-Cosmology-PIT/furry-parakeet}}) and those in the Python layer ({\tt pyimcom}\footnote{\url{https://github.com/Roman-HLIS-Cosmology-PIT/pyimcom}}).

\subsection{C routines}

The C routines are in the {\tt furry-parakeet} repository. These are functions that were difficult to build in an efficient way using standard vectorization with Python. They perform forward and transpose interpolation; the latter is needed for computing gradients of the cost function, and is most similar to the ``cloud-in-cell'' operator used in $N$-body simulations (e.g., \cite{Eastwood1974}).

\vskip 0.1in

\noindent{\tt bilinear\_interpolation}: This function takes inputs: 
\begin{list}{$\bullet$}{}
\item {\tt image}: the input image $\tilde I_{Bj}$.
\item {\tt rows}, {\tt cols}: the shape of $\tilde I_{Bj}$, and the intended shape of $X_{Ai}$.
\item {\tt g\_eff}: the gain image $G_{Bj}$ .
\item {\tt coords}: the coordinates of pixel $Bj$ in the coordinate system of image $A$. 
\item {\tt num\_coords}: the length $N$ of the coordinates vector.
\item {\tt interp\_data}: an image of zeroes in the same format as image $A$ ($X_{Ai}$); this is filled in in place.
\end{list}
The function carries out the operation:
\begin{equation}
    X_{Ai} = \sum_j w_{Ai;Bj} G_{Bj} \tilde I_{Bj}.
\end{equation}
Note that if pixel $i$ is ``out of bounds'' then $X_{Ai}$ is skipped, so  $X_{Ai}$ is initialized to zeros before the function call.

\vskip 0.1in

\noindent{\tt bilinear\_transpose}: This function takes inputs: 
\begin{list}{$\bullet$}{}
\item {\tt image}: the input gradient image $g_{Ai}$ .
\item {\tt rows}, {\tt cols}: the intended shape of $\gamma_{Bj}$.
\item {\tt coords}: the coordinates of pixel $Aj$ in the coordinate system of $B$. 
\item {\tt num\_coords}: the length $N$ of the coordinates vector.
\item {\tt original\_image}: an image of zeroes in the same format as image B ($I_{B}$); this is filled in in place.
\end{list}
The function carries out the operation:
\begin{equation}
    \gamma_{A\to B, j} = \sum_i w_{Ai;Bj} g_{Ai}.
\end{equation}
This skips a pixel $j$ if the total weight (denominator) is 0 and does not initialize the numerator internally, so $\gamma_{Aj}$ is initialized to zeros before the function call.

\subsection{Methods of the {\tt Parameters} class}

The following methods are part of the {\tt pyimcom.imDestripe. Parameters} class.

\vskip 0.1in

\noindent{\tt forward\_par}
(method of class {\tt parameters}): This method takes inputs: 
\begin{list}{$\bullet$}{}
\item {\tt j}: an index  to retrieve a certain parameter from {\tt params}, e.g., $j=0\to p_{A}$, $j=1\to p_{B}$.
\end{list}
It carries out the operation
\begin{equation}
    P_A = p_{A}{\textbf{1}}^{\rm T},
\end{equation}
returning a $N_{rows}\times N_{cols}$ matrix with values $p_{A\alpha}$ broadcast across rows.

\vskip 0.1in

\noindent{\tt subtract\_parameters} (method of class {\tt sca\_img}): This method takes inputs: 
\begin{list}{$\bullet$}{}
\item {\tt params}: a parameters object, containing the current working parameters vector.
\item {\tt k}: an index for {\tt params} to retrieve the right set for the {\tt sca\_img}.
\end{list}
First the function calls {\tt forward\_par} to make $P_A$. Then it carries out the operation
\begin{equation}
    \tilde{I}_B = I_B - P_B.
\end{equation}

\subsection{Functions in {\tt imDestripe.py}}

The following functions are in the {\tt pyimcom.imDestripe} module.

\vskip0.1in

\noindent{\tt transpose\_par}: This function takes inputs: 
\begin{list}{$\bullet$}{}
\item {\tt I}: an array, $N_{rows}\times N_{cols}$.
\end{list}
The function carries out the operation
\begin{equation}
    I_i= \sum_{j=0}^{N_{cols}}I_{ij}.
\end{equation}

\vskip 0.1in

\noindent{}{\tt make\_interpolated} (method of class {\tt sca\_img}): This function takes inputs: 
\begin{list}{$\bullet$}{}
\item{\tt sca\_A}: $\tilde{I}_A$, Image A with parameters $p_{A\alpha}$ subtracted and object and permanent mask applied.
\item{\tt ind}: the index of $I_A$ within the list of all SCAs in the mosaic.
\item{\tt params}: a parameters object with the current vector of all parameters for the SCAs, or None if first iteration.
\end{list}
This function carries out the operation: 
\begin{equation}
  \tilde J_{Ai}=\frac{\sum_{\text{ B in A}} X_{Ai}}{{\cal N}_{B|Ai} G_{Ai}}  
=\frac{\sum_{\text{ B in A}} \sum_j w_{Ai;Bj} G_{Bj} \tilde I_{Bj}}{{\cal N}_{B|Ai} G_{Ai}}  ,
\end{equation}
where $X_{Ai}$ is the image $A$ grid interpolation of a given image $B$, which we compute and add into $\tilde J_{Ai}$ only if $B$ overlaps $A$.

\vskip0.05in

\noindent{}The specific algorithm used is:
\begin{enumerate}
\item For the images $B$ that overlap image $A$: initialize $I_B$ and add the proper noise frame.
\begin{enumerate}
\item If {\tt params} is defined, call {\tt subtract\_parameters} to get $\tilde{I}_B$.
\item Apply the pixel mask to $\tilde{I}_B$.
\item Initialize an array of zeros for $\tilde{X}_A$ and $\tilde{J}_A$.
\item Apply {\tt interpolate\_image\_bilinear} to $\tilde{I}_B$ and its permanent mask onto the coordinate system of $I_A$, resulting in $\tilde{X}_A$ for the image and mask.
\item Add each $\tilde{X}_A$ to a cumulative interpolation image and a cumulative ${\cal N}_{B|Ai}$. 
\end{enumerate}
\item Once all SCAs overlapping $I_A$ have been accounted for, divide by ${\cal N}_{B|Ai}$ and $G_A$.
\end{enumerate}

\vskip 0.1in

\noindent{\tt cost\_function}: This function takes inputs: 
\begin{list}{$\bullet$}{}
\item {\tt p}: a vector of current de-striping parameters $p$. Initializes to an array of zeroes with shape $(N_{SCAs}, N_{rows})$.
\item {\tt f}: the functional form of the cost function, currently $f(x)=x^2$.
\end{list}

The function carries out the following algorithm:
\begin{enumerate}
\item For each SCA $A$:
\begin{enumerate}
    \item Initialize SCA image $I_A$, add noise, make (if iteration 1) or apply object mask.
    \item Get $\tilde I_A$ via {\tt subtract\_parameters} with $p_A$.
    \item Re-apply $I_A$ masks to set bright objects and bad pixels to zero.
    \item Get $\tilde J_A$ via $I_A$.{\tt make\_interpolated}, then apply the same image $A$ masks.
    \item Subtract $ \tilde I_A - \tilde J_A$ (for nonzero pixels) to obtain $\Psi_A$; update in $\Psi$ (the stack of $N_{SCA}$ $\Psi_A$s).
    \item Compute the local cost function $\epsilon_A = \sum_{ij}f(\Psi_{Aij})$.
\end{enumerate}
\item Finally, calculate the total cost by summing over SCAs: $\epsilon_T=\sum_A\epsilon_A$.
\end{enumerate}
The function returns:
\begin{equation}
\epsilon_T = \sum_A\sum_{ij}f(\Psi_{Aij}) = \sum_A\sum_{ij}f(\tilde I_{Aij} - \tilde J_{Aij})
\end{equation}
and
\begin{equation}
    \Psi = \tilde I - \tilde J.
\end{equation}

\vskip 0.1in

\noindent{\tt residual\_function}:

This function takes inputs: 
\begin{list}{$\bullet$}{}
\item {\tt $\Psi$}: the vector of difference images ($\tilde{I}_A-\tilde{J}_A$) for each SCA.
\item {\tt f'}: the functional form of the derivative of the cost function. This is currently $f'(x)=2x$ since the quadratic cost function is $f(x)=x^2$.
\end{list}
The function carries out the algorithm:
\begin{enumerate}
\item For each SCA $A$:
\begin{enumerate}
    \item Calculate the derivative $g_A=f'(\Psi_A)$.
    \item Sum up the value of $g_A$ across rows using {\tt transpose\_par} to get $t_{1A}$, an $N_{rows}$ vector representing the first term in $r_A = \nabla_{p_A}\epsilon_A$.
    \item Un-normalize the gradient image to prepare for transpose interpolation, $g_A/(G_A{\cal N}_{B|Ai})$.
    \item For overlapping SCA image ``B''s:
    \begin{enumerate}
        \item Initialize $I_B$.
        \item Initialize a transpose interpolated gradient $\gamma_B=0$.
        \item Use {\tt transpose\_interpolate} to get $\gamma_{A\to B, J}$ from $g_A$, and multiply by $G_B$.
        \item Sum up the values of $\gamma_{A\to B, j}$ over rows using {\tt transpose\_par} to get $t_{2B}$, an $N_{rows}$ vector.
        \item Add $t_{2B}$ to $r_B$.
    \end{enumerate}
    \item Subtract $t_{1A}$ from $r_A$.
\end{enumerate}
\item Return the full residual matrix $r$ ($N_{SCA}\times N_{rows}$).
\end{enumerate}

\noindent{}In summary, the function carries out the operation:
\begin{equation}
    r_{\alpha, A} = - \sum_i g_{Ai} \delta_{\alpha i} 
     + \sum_{B}\sum_j \gamma_{B \to A,j} G_{Bj} \delta_{\alpha j},
\end{equation}
summing up values across rows so that the resultant residual vector has $N_{rows}$ values per SCA. 
\vskip 0.05in
Expanding out the operations:
\begin{multline}
    r_{\alpha,A} = - \sum_i f'(\Psi_{Ai}) \delta_{\alpha, Ai} \\
     + \sum_{B} \Bigg[\sum_{j} \frac{ \sum_i w_{Ai;Bj} f'(\Psi_{Ai})}{G_{Ai}{\cal N}_{A|B}} G_{Bj} \delta_{\alpha, Bj}\Bigg].
\end{multline}




\vskip 0.1in

\noindent{\tt linear\_search}:
This function takes inputs: 
\begin{list}{$\bullet$}{}
\item {\tt p}: current parameters vector
\item {\tt d}: direction in which to search
\item {\tt r}: current gradient
\item {\tt f}: functional form for cost function
\item{\tt f'}: derivative of f
\end{list}

\noindent{}The function carries out the algorithm:
\begin{enumerate}
    \item Calculate the current $d\epsilon/d\alpha$ = $r\cdot d$,
    \item Find a test value for $\alpha$ based on curvature: $\alpha_{\text{test}}=-0.1 \frac{d\epsilon/ d\alpha}{\|\texttt{d}\|} $ and set the min and max $\alpha$ to $1\times10^{-4}$ and $10\times$ this value, respectively.
    \item Update a test set of parameters to $p_{\text{test}}=p+\alpha_{\text{max}}d$
    \item Calculate the cost function and residual for this test set of parameters.
    \item Directly calculate the direction step that minimizes the derivative of the cost function: $\alpha_{\text{new}} = \frac{-\alpha_{\text{max}}(d\cdot r)}{d\cdot(r_{\text{test}}-r)}$ .
    \item Update the parameters to $p_{\text{new}}=p+\alpha_{\text{new}}d$.
    \item Calculate the new cost function and residuals for this set of parameters.
\end{enumerate}
The function returns the new values of the parameters, the updated $\Psi$ for the new parameters set, and the corresponding cost and residual values.

\vskip 0.1in

\noindent{\tt conjugate\_gradient}:
This function takes inputs: 
\begin{list}{$\bullet$}{}
    \item {\tt p}: a parameters object, initially with all parameters set to zero.
    \item {\tt f, f'}: the cost function form and its derivative.
    \item {\tt tol}: the tolerance level; if the norm of the gradient reaches this value, say the solution has converged.
    \item {\tt max\_iter}: the maximum number of conjugate gradient iterations to go through before stopping, even if not converged.
    \item {\tt ov\_mat}: the overlap matrix, containing fractional overlap of SCAs in this mosaic.
\end{list}
The function carries out the process:
\begin{enumerate}
\item For each iteration $k<=\texttt{max\_iter}$:
\begin{enumerate}
    \item Calculate the initial cost $\epsilon_k$ and difference image $\Psi_k$ using {\tt cost\_function($p_k$,f)}
    \item\label{iter} Calculate the gradient $g_k$ using {\tt residual\_function} of $\Psi_k$
    \item Calculate $n_k$, the vector norm of $g_k$, and 
    \begin{enumerate}
        \item for iteration $k=0$: set $n=n_0$, tol$=n_0*$tol, and direction $d_0=-g_0$
        \item for all other iterations: calculate $\beta=|g_k|^2/|g_{k-1}|^2$, and set $d_k=-g_k+\beta_k d_{k-1}$
    \end{enumerate}
    \item Find the best possible new parameters $p_{k+1}$ and get the corresponding difference image $\Psi_{k+1}$ using {\tt linear\_search} in the direction $d_k$ with parameters $p_k$
    \item Make updates: $p_{k+1}\to p_k$, $\Psi_{k+1}\to\Psi_k$, $g_k\to g_{k-1}$, $d_k\to d_{k-1}$
    \item If convergence or maximum iterations are reached, finalize the parameters $p$. Otherwise,
    \item return to Item \ref{iter} and repeat the process until either convergence or maximum iterations.
\end{enumerate}
\end{enumerate}

\end{document}